\documentclass[useAMS,usenatbib]{mn2e}

\usepackage{natbib,graphicx,epsfig,rotating,color,verbatim}
\usepackage{colortbl}
\usepackage{footnote, threeparttable}
\definecolor{Gray}{gray}{0.9}

\newcommand{\myemail}{andrea.caliandro@ieec.uab.es}

\newcommand{\fermi}{{\it Fermi}}

\newcommand{\lsi}{LS~I~+61$^{\circ}$303}
\newcommand{\ls}{LS\,5039}
\newcommand{\hess}{HESS J0632+057}
\newcommand{\apjl}{ApJ}
\newcommand{\apj}{ApJ}
\newcommand{\aap}{A\&A}
\newcommand{\mnras}{MNRAS}          
\title[Searching for $\gamma$-ray emission from HESS J0632+057]{The missing GeV $\gamma$-ray binary:\\ Searching for HESS J0632+057 with \fermi-LAT}
\author[G.A. Caliandro, A.B. Hill et al.]{G.~A. Caliandro$^{1}$\thanks{Contact author: \myemail}, A.~B. Hill$^{2,3}$\thanks{Contact author: A.Hill@soton.ac.uk},
D.~F. Torres$^{1,4}$,
D. Hadasch$^{1}$,
P. Ray$^{5}$,
A. Abdo$^{6}$,
\newauthor J. W. T. Hessels$^{7,8}$,
A. Ridolfi$^{9}$,
A. Possenti$^{9}$,
M. Burgay$^{9}$,
N. Rea$^{1}$,
\newauthor P. H. T. Tam$^{10}$,
R. Dubois$^{2}$,
G. Dubus$^{11}$,  
T. Glanzman$^{2}$ and
T. Jogler$^{2}$
 \\
$^{1}$Institut de Ciencies de l'Espai (IEEC-CSIC), Campus UAB, 08193 Barcelona, Spain\\
$^{2}$W. W. Hansen Experimental Physics Laboratory, Kavli Institute for Particle Astrophysics and Cosmology, \\ Department of Physics and SLAC National Accelerator Laboratory, Stanford University, Stanford, CA 94305, USA\\ 
$^{3}$Faculty of Physical \& Applied Sciences, University of Southampton, Highfield, Southampton, SO17 1BJ, UK\\ 
$^{4}$Instituci\'o Catalana de Recerca i Estudis Avan\c{c}ats (ICREA), Barcelona, Spain\\
$^{5}$Space Science Division, Naval Research Laboratory, Washington, DC 20375, USA\\
$^{6}$Center for Earth Observing and Space Research, College of Science, George Mason University, Fairfax, VA 22030, USA.\\
  Resident at Naval Research Laboratory, Washington, DC 20375, USA.\\
$^{7}$ASTRON, the Netherlands Institute for Radio Astronomy, Postbus 2, 7990 AA, Dwingeloo, The Netherlands\\
$^{8}$Astronomical Institute "Anton Pannekoek," University of Amsterdam, Science Park 904, 1098 XH Amsterdam, The Netherlands\\
$^{9}$INAF Osservatorio Astronomico di Cagliari, Loc. Poggio dei Pini, 09012 Capoterra (CA), Italy\\
$^{10}$Institute of Astronomy and Department of Physics, National Tsing Hua University, Hsinchu 30013, Taiwan\\
$^{11}$Institut de Plan\'etologie et d'Astrophysique de Grenoble (IPAG), UJF-Grenoble 1 / CNRS-INSU, UMR 5274, \\ Grenoble, F-38041, France\\
}

\begin{document}

\date{Accepted 201X XXXXXX XX. Received 201Y YYYYYYY YY; in original form 2012 December ZZ}

\pagerange{\pageref{firstpage}--\pageref{lastpage}} \pubyear{2002}

\maketitle

\label{firstpage}

\begin{abstract}
The very high energy (VHE; $>$100 GeV) source HESS J0632+057 has been recently confirmed as a $\gamma$-ray binary, a subclass of the high mass X-ray binary (HMXB) population, through the detection of an orbital period of 321 days.  
We performed a deep search for the emission of HESS J0632+057 in the GeV energy range using data from the \fermi\ Large Area Telescope (LAT). The analysis was challenging due to the source being located in close proximity to the bright $\gamma$-ray pulsar PSR J0633+0632 and lying in a crowded region of the Galactic plane where there is prominent diffuse emission. We formulated a Bayesian block algorithm adapted to work with weighted photon counts, in order to define the off-pulse phases of PSR J0633+0632.
A detailed spectral-spatial model of a 5$^{\circ}$ circular region centred on the known location of \hess\ was generated to accurately model the LAT data.  No significant emission from the location of HESS J0632+057 was detected in the 0.1--100 GeV energy range integrating over $\sim$3.5 years of data; with a 95\% flux upper limit of F$_{0.1-100~{\rm GeV}}$ $< 3 \times 10^{-8}$ ph cm$^{-2}$ s$^{-1}$.  A search for emission over different phases of the orbit also yielded no significant detection.  A search for source emission on shorter timescales (days--months) did not yield any significant detections.
We also report the results of a search for radio pulsations using the 100-m Green Bank Telescope (GBT).  
No periodic signals or individual dispersed bursts of a likely astronomical origin were detected.
We estimated the flux density limit of $< 90/40\,\mu$Jy at 2/9\,GHz.
The LAT flux upper limits combined with the detection of \hess\ in the 136--400 TeV energy band by the MAGIC collaboration imply that the VHE spectrum must turn over at energies $<$136 GeV placing constraints on any theoretical models invoked to explain the $\gamma$-ray emission.
\end{abstract}


\section{Introduction}

The population of $\gamma$-ray binaries is a subclass of the high mass X-ray binary (HMXB) population; they are binary systems comprised of a high mass ($>10$ M$_\odot$) star and a compact object whose spectral energy distribution (SED) peaks above 1 MeV.
Indeed, they exhibit emission at high energies (HE; 0.1--100\, GeV) and/or very high-energies (VHE; $>$100 GeV).  The non-thermal high energy radiation from these objects indicates that they are sites of natural particle acceleration.  At present 
only a handful of confirmed $\gamma$-ray binaries are known:
\begin{itemize}
\item [-] \lsi\ \citep{2006Sci...312.1771A, 2008ApJ...679.1427A, 2009ApJ...701L.123A,2012ApJ...749...54H}
\item [-] \ls\ \citep{2005Sci...309..746A,2009ApJ...706L..56A,2012ApJ...749...54H}
\item [-] PSR B1259$-$63 \citep{2005A&A...442....1A,B1259,B1259Tam}
\item [-] \hess\  \citep{hess_discovery, hinton2009, Bongiorno2011}
%\item [-] Cyg X$-$3 \citep{2009Sci...326.1512F, 2009Natur.462..620T, 2012MNRAS.421.2947C}
\item [-] 1FGL J1018.6$-$5856 \citep{1FGLbinary_science}
\end{itemize}

Cyg X$-$3 \citep{2009Sci...326.1512F, 2009Natur.462..620T, 2012MNRAS.421.2947C} and Cyg X$-$1 \citep{2007ApJ...665L..51A,2010ApJ...712L..10S} show $\gamma$-ray flaring activity, but their SEDs are peaked in X-rays. 
So they do not belong to the population of the $\gamma$-ray binaries.
% \noindent  reported the first experimental evidence of  as a flaring TeV and GeV source, however these reports are of low statistical significance and have not been confirmed by other instruments. 

\lsi, \ls\ and PSR B1259$-$63 have all been detected at both GeV and TeV energies;  1FGL J1018.6$-$585 was discovered at GeV energies solely with data from the
\textit{Fermi Gamma-ray Space Telescope} Large Area Telescope 
(\fermi-LAT) and has recently been reported to have a possible TeV counterpart, HESS J1018$-$589 \citep{2012arXiv1203.3215H}.

\hess\ was the first $\gamma$-ray binary to be discovered at VHE prior to its detection in other wavebands; the H.E.S.S. collaboration reported an unidentified point-like VHE source lying close to the Galactic plane \citep{hess_discovery}.  It was noted that a potential optical counterpart to the source was a massive emission-line star, MWC 148, of spectral type B0pe.  These facts led to the hypothesis that the source may be a new $\gamma$-ray binary, a nature that was confirmed by the discovery of an orbital period of $321 \pm 5$ days detected in the X-ray counterpart by \citet{Bongiorno2011} using data from a long-term \textit{Swift}-XRT monitoring campaign.  \citet{casares2012} determined the ephemeris and the orbital parameters of the system using the X-ray orbital period and through optical spectroscopy of MWC 148. They found the binary to be in a highly eccentric orbit ($e = 0.83 \pm 0.08$).

Following the discovery of \citet{hess_discovery} observations made by the VERITAS collaboration 
reported upper limits at the 99\% confidence level that are $~$2.4 times lower than the flux found by the H.E.S.S. collaboration, implying a variable flux behaviour of the source at VHE \citep{veritas_ul}.  \hess\ was detected again at  VHE by MAGIC and VERITAS in February 2011 when TeV observations were simultaneously taken with an outburst in X-rays \citep{falcone2011,veritas_icrc_2011,magic2012}.  In the X-ray regime \cite{hinton2009} identified the X-ray counterpart of \hess\,  XMMU J063259.3+054801 from \textit{XMM}-Newton observations.  The X-ray counterpart exhibits a hard spectrum with spectral index $\Gamma = 1.24 \pm 0.04$ that is observed to soften from 1.2 to 1.6 with rising flux \citep{reatorres2011}.  The source exhibits flux variability on timescales of hours. 

Similar X-ray characteristics together with a softer spectrum at TeV energies are also seen in other $\gamma$-ray binaries e.g. \lsi\ and \ls\ \citep{sidoli2006,esposito2007,kishishita2009,rea2010,rea2011}. \citet{hinton2009} explain the observed emission and variability as synchrotron and inverse Compton emission from a population of relativistic electrons in a region within a few AU of the star.  Similar to \lsi, the peak X-ray emission occurs $\sim$0.3 orbital phases after the periastron passage.

In the radio regime \citet{skilton2009} detected a point-like, variable radio source at the position of MCW 148 with the Giant Metrewave Radio Telescope (GMRT) and the Very Large Array (VLA) at 1.28 and 5\,GHz, respectively. \citet{moldon2011} observed \hess\ with the European VLBI Network (EVN) at 1.6\,GHz in two epochs: during the 2011 January/February X-ray outburst and 30 days later. In the first epoch the source appears point-like, whereas in the second it appears extended with a projected size of $\sim$75\,AU assuming a distance to the system of 1.5\,kpc. The brightness temperature of 2$\times$10$^{6}$\,K at 1.6\,GHz 
together with the average spectral index of $\sim$\mbox{0.6} is indicative of non-thermal synchrotron radiation producing the radio emission \citep{skilton2009,moldon2011}.  The projected displacement of the peak of the radio emission in the 30 day gap between observations is 21\,AU. The radio morphology and the size of \hess\ are similar to those observed in the well-established $\gamma$-ray binaries PSR B1259$-$63, \ls\ and \lsi.

Within the small class of known $\gamma$-ray binaries the nature of the compact source is known only for one system: PSR B1259$-$63 contains a 48\,ms radio pulsar \citep{2004MNRAS.351..599W}. 
%Cyg X$-$3 is clearly identified as a microquasar system known to harbour powerful jets and may be host to a black hole \citep[see e.g.][]{2010ApJ...718..488S}. 
The nature of the compact objects in \ls, \lsi\, 1FGL J1018.6$-$5856 and \hess\ is still unknown \citep{hill2010}. \cite{reatorres2011} conducted a \textit{Chandra} observing campaign during the high state of X-ray and TeV activity of \hess\ in 2011 February. Their timing analysis did not find pulsations in the X-ray data and they derived 3$\sigma$ upper limits on the X-ray pulsed fraction of $\sim$30\%, similar to those derived for \lsi\ and \ls\ \citep{rea2010,rea2011}.

Presented here are the results of a detailed analysis of $\sim$43 months of \fermi\ LAT survey observations of \hess. We describe the procedures implemented to analyse this complex field and the results of searching for persistent, orbitally variable and transient high-energy emission from \hess. 
A search for radio pulsations using the 100-m Green Bank Telescope (GBT) is also reported.

\section{Fermi LAT observations}
The Large Area Telescope (LAT) is the primary instrument onboard \fermi; it is a pair conversion telescope sensitive to photons with energies from $\sim$20\,MeV to more than
300\,GeV.  The direction of an incident photon is derived by tracking the electron-positron pair in a high-resolution converter tracker, and the energy of the pair is measured with a hodoscopic CsI(Tl) crystal
calorimeter. The \fermi-LAT has a peak on-axis effective area of 8000 cm$^2$, a 2.4\,sr field of view, and an angular resolution of $\sim$0.6$^{\circ}$ at 1\,GeV (for events converting in the front
section of the tracker).  Furthermore, an anti-coincidence detector vetoes the background of charged particles \citep{Atwood2009}.

The observatory operates principally in survey mode; the telescope is rocked north and south on alternate orbits to provide more uniform coverage so that every part of the sky is observed for $\sim$30 minutes every 3 hours.  The analysis presented here makes use of all data taken from 2008 August 4 until 2012 March 3.  

\subsection{Data reduction}
The \fermi\ \textsc{Science Tools v09r28p00} package was used to analyse the `source' event class data. 
%The analysis used the `{\tt source}'  photon class events from the `{\tt P7.6\_P130\_BASE}' dataset. 
All events 
within a circular region of interest (ROI) of 15$^{\circ}$ radius centred on the known location of the massive star MWC 148 associated with HESS J0632+057 (R.A.(J2000) = 98\fdg247, Dec(J2000) = 5\fdg800) and in the energy range 0.1--100 GeV were extracted.  The good time intervals are defined such that the ROI does not cross the $\gamma$-ray-bright Earth limb (defined at 100$^{\circ}$ from the zenith angle), and that the source is always inside the LAT field of view, namely within a cone angle of 66$^{\circ}$.  The `{\tt P7SOURCE\_V6}' instrument response functions (IRFs) are used throughout the analysis. 

When constructing spectral-spatial models of the ROI the models used for the Galactic diffuse emission (gal\_2yearp7v6\_v0.fits) and isotropic backgrounds (iso\_p7v6source.txt) were those currently recommended by the LAT team\footnote{Descriptions of the models are available from the FSSC: http://fermi.gsfc.nasa.gov/} and were used by the \fermi-LAT collaboration to build the Second \fermi-LAT Source Catalog \citep[hereafter 2FGL,][]{2FGL}.

\section{Data analysis: The $\gamma$-ray sky around HESS J0632+057}

\begin{figure}
\center
\includegraphics[width=0.5\textwidth]{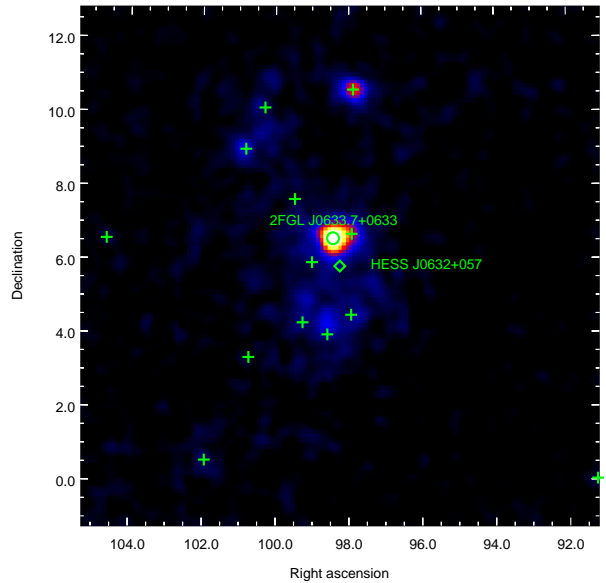}
\caption{Background subtracted count map above 1 GeV of the $10^{\circ} \times 10^{\circ}$ sky region centered on HESS J0632+057. The pixel size is 0\fdg1$\times$0\fdg1 and a Gaussian smoothing with 3 pixel kernel radius is applied. The 2FGL sources are labeled with green crosses, HESS J0632+057 with a diamond and PSR J0633+0632 with the circle.}
\label{Cmap}
\end{figure}

\hess\ lies at the Galactic latitude $b=-1\fdg44$, in the active region of the Monoceros Loop. An image of this region in the $\gamma$-ray band is shown in Figure \ref{Cmap}.  This is a background subtracted count map of the $10^{\circ} \times 10^{\circ}$ region around \hess. It is generated by selecting events with energies greater than 1 GeV, and subtracting the counts due to the Galactic diffuse emission and the isotropic emission as calculated using the respective models with the \textsc{Science Tool} {\tt gtmodel}.  In Figure \ref{Cmap} it is evident that the region is dominated by the $\gamma$-ray emission of PSR J0633+0632 (2FGL J0633.7+0633), which is only 0\fdg78 away from \hess.

\subsection{PSR J0633+0632}

PSR J0633+0632 is a bright, radio-quiet $\gamma$-ray pulsar discovered in the first six months of the \fermi\, mission \citep{16blind}.  It has a flux ($>$100 MeV) of $(8.4 \pm 1.2) \times 10^{-8}$ ph cm$^{-2}$ s$^{-1}$, and the pulse profile has the characteristic two narrow peaks separated by $\sim 0.5$ in phase \citep{1PC}. 

\begin{figure}
\center
\includegraphics[width=0.45\textwidth]{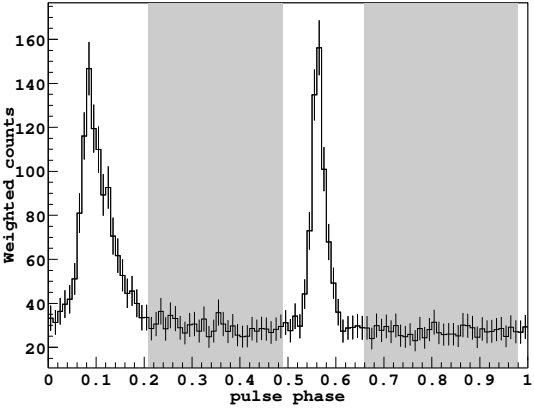}
\caption{Weighted pulse profile of PSR J0633+0632 using $\sim 3.5$ years of \fermi-LAT data. 
%The weight associated to each event count corresponds to the probability that the event is emitted by the pulsar, rather than by a nearby source or by the diffuse emission. 
The off-pulse, and the intra-peak phases (shaded regions) are defined by means of the Bayesian block algorithm: (0.66 - 0.98) and (0.21 - 0.49), respectively.}
\label{PSRJ0633}
\end{figure}

The pulse profile obtained with the dataset analysed in this paper is shown in Figure \ref{PSRJ0633}.  To coherently assign a phase to each photon over such a long observation period ($\sim3.5$ years) requires an up-to-date ephemeris for the pulsar.  As PSR J0633+0632 is radio-quiet, the new ephemeris has been calculated using the \fermi-LAT data with the method described by \citet{PRay}. A phase was associated to each event using the \fermi\, plug-in of the \textsc{tempo2} software package \citep{TEMPO2}. A weight was also assigned to each event, corresponding to the probability that the gamma ray is emitted by the pulsar, rather than by a nearby source or by the diffuse emission \citep{2011ApJ...732...38K}; weights are calculated using the \textsc{Science Tool} {\tt gtsrcprob}. The spectral models of the pulsar and the nearby sources adopted to calculate the weights are those from the 2FGL catalogue. Finally, the pulse profile in Figure \ref{PSRJ0633} was obtained by summing the weights in each phase bin for the events within 2\degr\ of the pulsar position. 

To minimise contamination from the strong emission of PSR J0633+0632 in the analysis of \hess\  the pulsar emission was `gated'\footnote{For a detailed example of pulsar gating see the FSSC tutorial: http://fermi.gsfc.nasa.gov/ssc/data/analysis/scitools/\\pulsar\_gating\_tutorial.html} by only using events which 
occurred outside the pulse peaks, during phases 0.66--0.98 and 0.21--0.49 (hereafter out-of-peak phases).  
The off-pulse and intra-peak phases of PSR J0633+0632 were defined by applying a Bayesian block algorithm adapted for weighted-counts light curves, described in the next section.

\subsection{Off-pulse and intra-peak definition}
We found that an efficient way to define the 
off-pulse phase range is by subdividing the pulse profile
into `Bayesian blocks', and choosing the block with the lowest count rate.
The intra-peak phase range is instead the lowest block between the two peaks.  
The Bayesian blocks algorithm was proposed by \citet{Scargle}: the algorithm 
defines the phase ranges of constant rate, called Bayesian blocks,
on the basis of the likelihood ratio test applied to a Poissonian distribution. This is perfect for count light curves, but not for weighted pulse profiles, as in this case the weights are not integer values, and they do not follow a Poissonian distribution. Furthermore, the pulse profile is periodic, meaning that it does not have a well defined start and end.

In order to manage weighted counts, in our method the Bayesian blocks are defined using the $\chi^2$ ratio test.   Specifically, given a light curve of weighted counts, the null hypothesis is that the light curve is flat, and it is fitted by a single block. The associated $\widetilde{\chi}^2$ value for a single block is defined as:
\begin{equation}
\widetilde{\chi}^2_{\rm 1bl.} = \frac{1}{N-1}\sum_i \frac{(B_{w\_i} - M_w)^2}{M_w}
\label{chi1}
\end{equation}
where $B_{w\_i}$ are the weighted source counts in the $i$-th bin, 
and $M_w$ are the average weighted counts all along the light curve. $N$ is the total number of bins.
The test hypothesis is that two blocks fit better than a single block. The $\widetilde{\chi}^2$ of the two blocks hypothesis is
\begin{equation}
\widetilde{\chi}^2_{\rm 2bl.} = \frac{1}{N-2} \left[ \sum_i \frac{(B_{w1\_i} - M_{w1})^2}{M_{w1}} + \sum_j \frac{(B_{w2\_j} - M_{w2})^2}{M_{w2}}\right] 
\label{chi2}
\end{equation}
where $M_{w1}$, $B_{w1\_i}$ and $M_{w2}$, $B_{w2\_j}$ are the average weighted counts, and the weighted counts in the $i$-th and $j$-th bin of the first and second block, respectively.
Finally, the significance of the improvement given by two blocks rather than a single block is calculated by means of the F-test. 

%If two blocks fit the data significantly better than a single block then the division point between the blocks is recorded and each of the new blocks are tested for further subdivision.  The process continues until further subdivision is not significant or the block length reaches a predefined minimum length.   

If two blocks fit the data significantly better than a single block then the division point between the blocks is recorded and 
the first of the new blocks (on the left) is tested for further subdivision. The process continues until further subdivision is not significant or the block length reaches a predefined minimum length. 
%In this way, the shorter most left significant block is defined. The subdivition restarts considering all the rest of the light curve on the right.
In this way, the shortest, furthest-left significant block
is defined. The subdivision restarts considering the remaining light
curve on the right.

The issue relating to the periodicity of the pulse profile is solved by selecting the starting (and ending) point of the light curve to be the phase of the highest peak. This is the choice that least affects the definition of the off-pulse region. 

The algorithm has three parameters: the number of bins of the pulse profile (100), the minimum possible length of a block (3 bins), and the threshold on the two blocks significance (1.5$\sigma$). The values in parenthesis are the settings used for PSR J0633+0632. We tested this algorithm on all the $\gamma$-ray pulsars detected so far with the \fermi-LAT, finding that these settings give good results for all the bright pulsars.

\subsection{Out-of-peak analysis}

\begin{figure*}
\center
\includegraphics[width=0.45\textwidth]{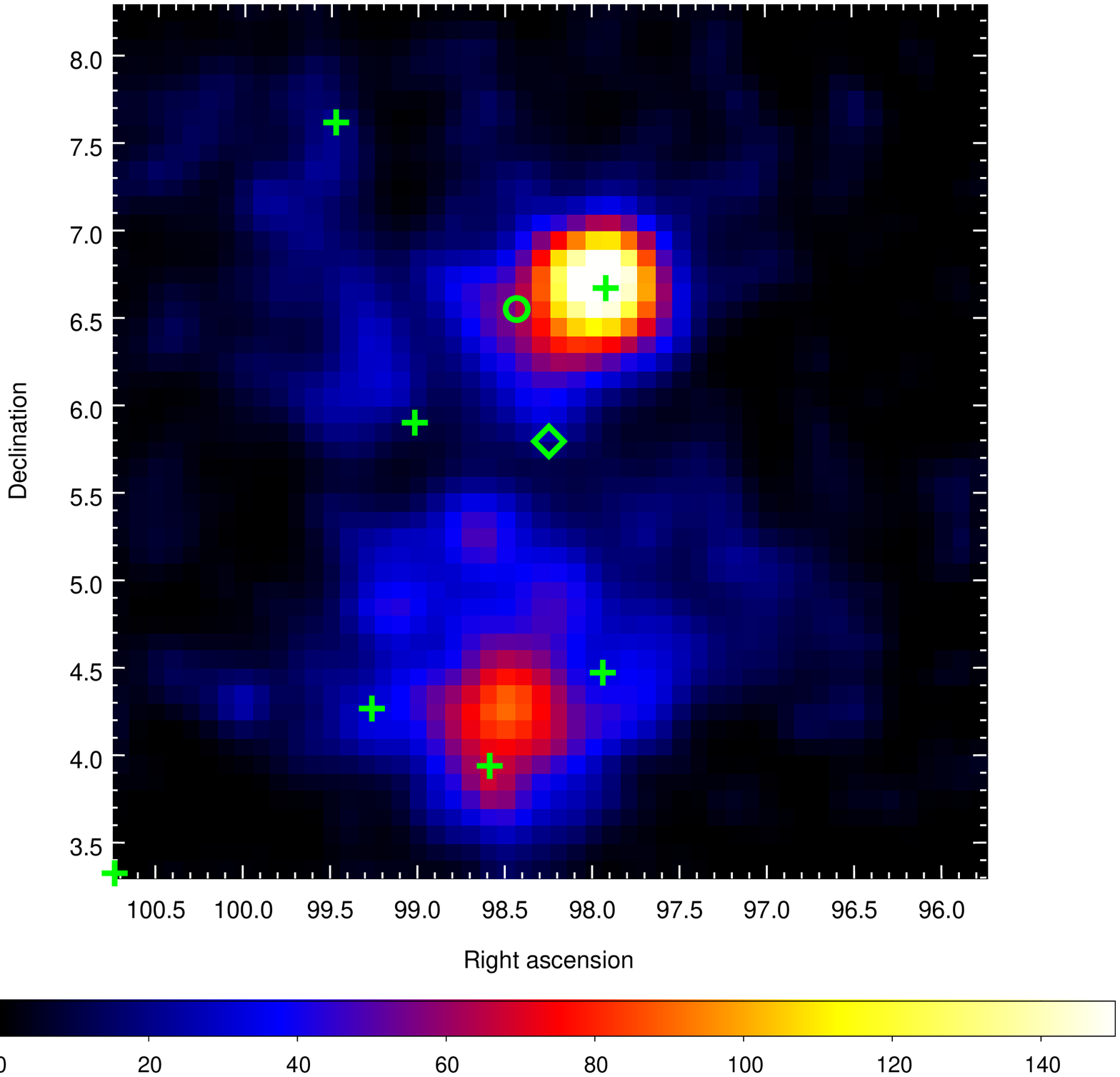}
\hspace{0.5 cm}%
\includegraphics[width=0.45\textwidth]{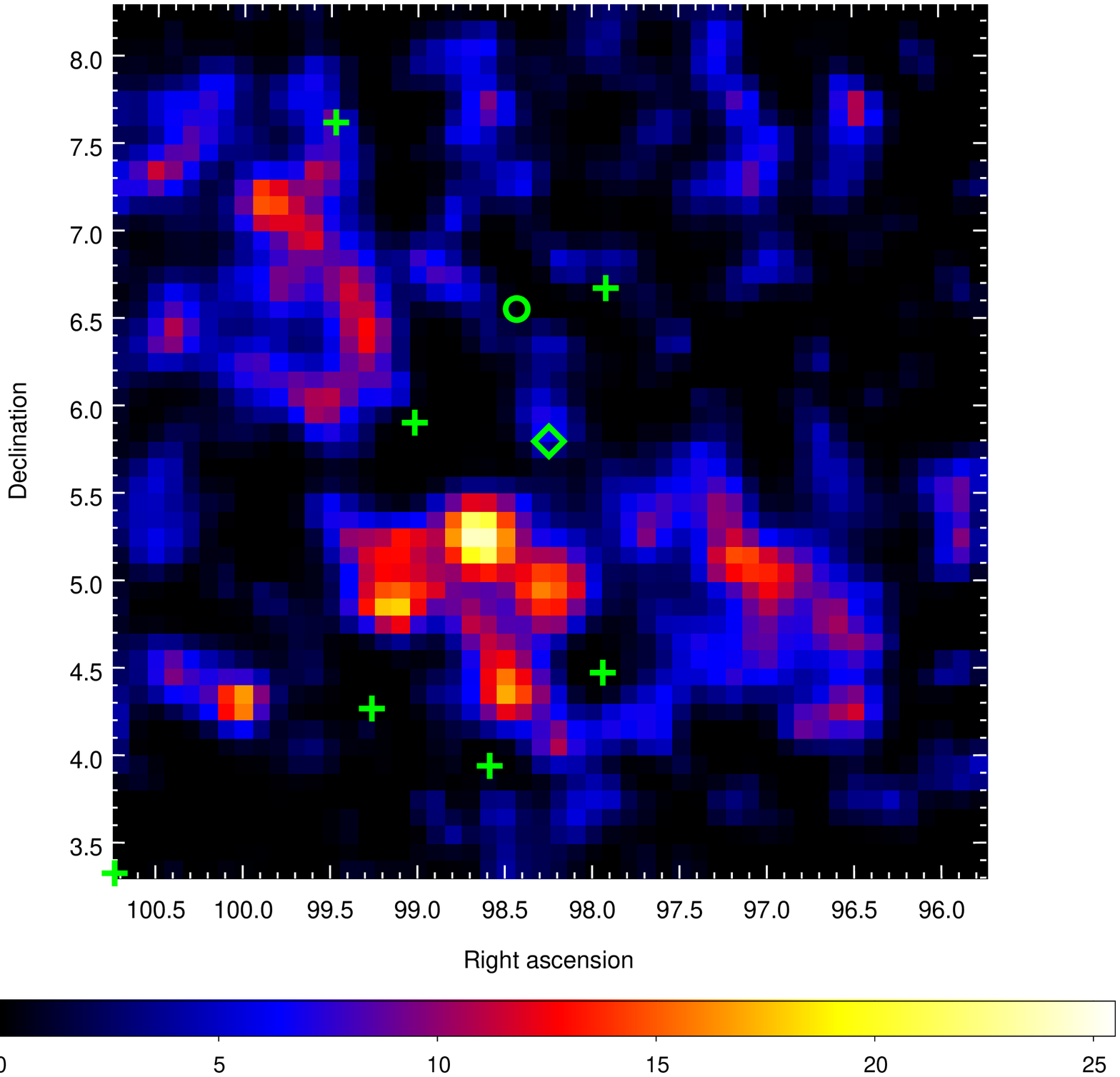}
\caption{TS map (\textit{left}), and residual TS map (\textit{right}; 2FGL sources are included in the model) of the $5^{\circ} \times 5^{\circ}$ sky region centered on HESS J0632+057. The maps are calculated for $E > 300$ MeV using photons in the off-pulse and intra-peak phases of PSR J0633+0632. The 2FGL sources are labeled with green crosses, HESS J0632+057 with a diamond, and the position of PSR J0633+0632 with the circle.
\label{TSmaps}}
\end{figure*}

Having restricted the events to those selected from within the out-of-peak phases of PSR J0633+0632 (shaded phase ranges in Figure \ref{PSRJ0633}), the analysis of the region around HESS J0632+057 was performed by means of the binned maximum-likelihood method \citep{Mattox1996}.  Two different tools were used to perform the spatial and spectral analysis: \textsc{Science Tool} {\tt gtlike} and \textsc{pointlike}. 
{\tt gtlike} is the official tool supplied with the publicly available \textsc{Science Tools} to fit spectral models to LAT data. 
\textsc{pointlike} is an alternative binned likelihood implementation, optimised to localise the source and to characterise its extension. It operates quickly on long datasets making it better suited for iteratively constructing source models of a region. Results from \textsc{pointlike} have been extensively compared with {\tt gtlike} and found to be consistent \citep[see e.g.,][]{Kerr2011,Joshua2012}.
The likelihood-ratio test statistic (TS) is employed to evaluate the significance of the $\gamma$-ray flux of a test source. The TS value is defined to be:
\begin{equation}
TS = -2\Delta\log{\mathcal{L}} = -2\log{\frac{\mathcal{L}(Null)}{\mathcal{L}(Test)}} 
\end{equation}
where $\mathcal{L}$ is the likelihood function maximised by adjusting all parameters of the model with and without the test source (test and null hypothesis, respectively).  A TS = 25 corresponds to a significance of just over 4$\sigma$ when the testing source is modelled with a simple power law \citep{2FGL}.

Figure \ref{TSmaps} shows two TS maps of the $5^{\circ} \times 5^{\circ}$ region centred on \hess. They are generated by testing for an additional point-like source located at the centre of each pixel (of size $0.1^{\circ} \times 0.1^{\circ}$). 
In the left panel of Figure \ref{TSmaps} is plotted a diffuse subtracted TS map, meaning that the spectral-spatial model for the null hypothesis includes only the Galactic and isotropic emission models.
In contrast, the right panel shows the residual TS map when the spectral-spatial model also includes the 2FGL sources within a radius of 20$^{\circ}$ from the ROI centre, in addition to the diffuse models.  The 2FGL sources within 5$^{\circ}$ of HESS J0632+057 (11 sources) are modelled with their flux normalizations
allowed to vary, while the other 45 sources had all their parameters fixed to the values given in the 2FGL catalog.  The spectral model of the isotropic diffuse emission had free normalization
and the spectral model of the Galactic diffuse emission is scaled by a simple power-law with normalisation and index free to account for any local variations in the diffuse emission.  In the final fit of the data the fitted value for the power-law index is consistent with 0 within the fit error, indicating that no adjustment to the local Galactic diffuse spectral shape was required. 

HESS J0632+057 lies in a region of high Galactic diffuse emission.
%In constructing the official LAT Galactic diffuse model 0.3--20 GeV data were used to fit the model, which was then extrapolated and renormalized to span the 50 MeV -- 100 GeV energy range\footnote{A description of the construction of the Galactic diffuse model is available from the FSSC:\\http://fermi.gsfc.nasa.gov/ssc/data/access/lat/\\ring\_for\_FSSC\_final4.pdf}. 
Since at low energies the diffuse emission is stronger, as well as its uncertainties, the following analysis was restricted to using data at $E>300$ MeV. 

The TS maps in Figure \ref{TSmaps} show that the region around HESS J0632+057 is not well modelled solely by the sources from the 2FGL catalogue. This is not unexpected considering that the 2FGL catalog was constructed with 2 years of \fermi-LAT data, while in this paper the dataset is almost doubled ($\sim 3.5$ years).  Searching for emission from \hess\ requires an accurate modeling of the entire region, and hence an improved source model of the region is required.  The ROI source model was iteratively improved by adding a new point source at the location of the highest residual TS and then re-calculating the residual TS map (see Sec. 3.3.3). Initially the 3$^\circ$ region around \hess\ was fine tuned before extending the radius to 5$^\circ$.  A detailed description of the process and results are given in the following sections. 

\begin{table*}
\centering
\caption{The sources included in the final spectral-spatial model within a radius of 5$^{\circ}$ of HESS J0632+057. The table is split into four sections: 2FGL catalogue sources; 2FGL sources relocalized in our procedure (named Reloc instead of 2FGL); new sources introduced to model the TS residuals; and \hess.  The Flux is for $E>300$ MeV in units of $10^{-9} $ cm$^{-2}$ s$^{-1}$. The flux of HESS J0632+057 is not reported here, because its TS value indicates that the source is not detected in the data.
}\label{Table1}
\begin{threeparttable}[b]
\begin{tabular}{@{}lccccc@{}}
\hline
Source Name &  R.A. (J2000) & Dec. (J2000) &  Spectral shape &  Flux & TS \\
\hline
2FGL J0631.5+1035 &  97.88 &  10.60 &  PLExpCutoff & 17.4 $\pm$ 1.3  &  350 \\
2FGL J0631.6+0640 &  97.92 &  6.68 & LogParabola  & 9.3 $\pm$ 1.2  &  140 \\
2FGL J0633.7+0633$^{\dagger}$ & 98.432  &  6.556 &  PowerLaw &  9.6 $\pm$ 2.2 &  37 \\
2FGL J0641.1+1006c & 100.29  & 10.10  & LogParabola  &  12.5 $\pm$ 1.8 &  71 \\
2FGL J0642.9+0319 & 100.73  & 3.33  &  LogParabola & 9.9 $\pm$ 2.0  &  30 \\
2FGL J0643.2+0858 &  100.82 &  8.98 & LogParabola  & 17.0 $\pm$ 1.6  &  170 \\

\hline %RELOC
Reloc J0631.7+0428$^{\ddag}$ &  97.70 & 4.80  &  LogParabola &  9.1 $\pm$ 1.7 &  45 \\
Reloc J0634.3+0356c$^{\ddag \diamond}$ & 98.52  &  4.27 &  PowerLaw & 13.2 $\pm$ 2.4  &  64 \\
Reloc J0636.0+0554$^{\ddag}$ & 99.28  & 6.10  &  LogParabola &  11.3 $\pm$ 1.9 &  54 \\
Reloc J0637.0+0416c & 99.34  & 4.45  & LogParabola  &  8.1 $\pm$ 2.0 &  27 \\

\hline %New
Fermi J0642.4+0648$^{\star}$  & 101.60  &  6.80 &  LogParabola &  8.9 $\pm$ 1.4 &  53 \\
New1$^{\diamond}$  & 98.54  &  3.74 &  PowerLaw & 10.1 $\pm$ 2.2  &  36 \\
New2$^{\S}$ & 99.65  & 7.17  &  PowerLaw &  8.7 $\pm$ 1.8 &  34 \\
New4$^{\star}$  & 96.55  &  8.80 &  PowerLaw &  4.8 $\pm$ 1.2 &  22 \\
New5$^{\star}$  & 99.70  & 9.47  &  PowerLaw &  6.5 $\pm$ 1.6 &  24 \\
New6 &  98.66 & 5.27  &  PowerLaw &  7.2 $\pm$ 1.8 &  28 \\
New7 &  100.04 &  4.32 & PowerLaw  & 6.1 $\pm$ 1.6  &  23 \\
New8$^{\star}$  & 97.05  & 5.10  &  PowerLaw &  6.2 $\pm$ 1.9 &  25 \\
\hline
\rowcolor{Gray}
HESS J0632+057 &  98.247 & 5.800  &  PowerLaw &  -- &  4.7 \\
\hline
\end{tabular}
\begin{tablenotes}
\item
$^{\dagger}$ PSR J0633+0632\\
$^{\star}$ Also detected in the $>100$ MeV iterative analysis.\\
$^{\ddag}$ Source position shifted from that found in the 2FGL catalogue beyond its 95\% confidence radius. \\
$^{\diamond}$ The original 2FGL J0634.3+0356c source has been resolved into two separate sources in this analysis.\\ %the brighter of the two retains the 2FGL catalogue name.\\
$^{\S}$ $\sim$0\fdg5 from the catalog position of 2FGL J0637.8+0737 which was removed from our initial source model (see text for details).\\
\end{tablenotes}
\end{threeparttable}
\end{table*}

\subsubsection{TS calculation of 2FGL sources}\label{stage1}
In the spectral-spatial model described in the previous section an additional point source was added at the position of HESS J0632+057. Its spectrum is modelled with a simple power-law, with flux and spectral index parameters set free. The TS values of all the 2FGL sources within $3^{\circ}$ of HESS J0632+057 were calculated from a maximum likelihood fit with the {\tt gtlike} tool.

With the exception of PSR J0633+0632 and HESS J0633+057, only one source, 2FGL J0637.8+0737, had a TS $< 50$. Four sources (2FGL J0636.0+0554, 2FGL J0631.7+0428, 2FGL J0634.3+0356c, 2FGL J0637.0+0416c) were found with TS values between 50 and 100 and only 2FGL J0631.6+0640 had a TS $> 100$.  As an initial step the source with TS $< 50$ was deleted from the spectral-spatial model and those sources with $50<$TS$<100$ were re-localised to their optimum positions using \textsc{pointlike}.  

%Using this new model the region was fit with the {\tt gtlike} tool allowing all source spectral parameters free in the fit.  

\subsubsection{Source spectrum modelling}
The residual off-pulse emission of PSR J0633+0632 is very weak, with TS $= 23$.  As a consequence the spectral model was changed from the super-exponential power-law, typical of $\gamma$-ray pulsars, to a simple power-law.

After this change we started the source spectrum modelling. In order to reach a stable solution for all the sources in the region, we first fit the brightest sources, keeping frozen the other sources to their nominal spectra. Then, we set free weaker and weaker sources, until in the last fit all sources had free spectral parameters.  In practice, this consisted of three steps. First, a maximum likelihood fit was performed with {\tt gtlike} setting free all the spectral parameters of the sources with TS $>90$ within 3$^{\circ}$ of HESS J0632+057. In the second step the fit was repeated setting free also the spectral parameters of the sources with TS between 50 and 90. Finally, in the third step the spectral parameters of all the sources were set free.

This established the best model of the region on the basis of the 2FGL catalogue source population from which 
additional, newly-found sources could be assessed.

\subsubsection{Adding new sources}
An iterative procedure was implemented to model excesses in the residual TS map as additional point sources in the model.  The process consisted of the following stages:
\begin{itemize}
 \item[-] Calculate the residual TS map using the current spectral-spatial model.
 \item[-] Add an additional point source to the model at the location of the highest excess in the TS map, modeled by a simple power law.
 \item[-] Fit the best position of the additional source with the \textsc{pointlike} tool.
 \item[-] Check the curvature of the spectrum of the additional source. If curvature is indicated, substitute the initial power-law model with a log parabola. 
 \item[-] Re-fit with {\tt gtlike} the spectra of all the sources in the region, setting free also the spectral parameters of the additional source.
\end{itemize}

The procedure was repeated until the maximum value within the residual TS map was 
less than 15, below which the procedure would effectively be fitting only statistical noise.  
Five additional point sources were added within 3$^{\circ}$ of \hess. Three more point sources were added when the procedure was repeated enlarging the radius to 5$^{\circ}$.
%Eight additional point sources were required to reduce the maximum TS map residual to be $<15$.  
Table~\ref{Table1} summarises the results obtained by modelling the region around \hess; listed are all the sources included in the final spectral-spatial model within a radius of 5$^{\circ}$ from HESS J0632+057.

Of the eight additional sources one source, `New1', may be a result of resolving 2FGL J0634.3+0356c into two sources, as the nominal position of 2FGL J0634.3+0356c was shifted when it was re-localised in our fitting procedure.  Note that this 2FGL catalogue source was denoted as a `c' source and hence is found in a region of bright and/or possibly incorrectly modelled diffuse emission \citep{2FGL}.  It is worth noting that `New2' is 29\arcmin\ from 2FGL J0637.8+0737 (the source deleted from the model in the initial stage, see \S~\ref{stage1}) although it lies outside the 95\% error radius for this catalogue source.

The majority of the additional sources are very weak.  Due to this and the challenge of modeling the Galactic diffuse emission it is very difficult to definitively characterize them as new $\gamma$-ray sources with any confidence rather than statistical fluctuations in the data or inaccurately modelled regions of the Galactic diffuse emission.  Consequently, we only report one new $\gamma$-ray source, Fermi J0642.4+0648, that is detected at $>7\sigma$ at a location of RA(J2000)=101\fdg60, Dec.(J2000)=6\fdg80 with a 95\% error radius of $\sim$7.8\arcmin. The source spectrum is modeled by a log parabola shape:
\begin{equation}
\frac{dN}{dE} = N_0 \left( \frac{E}{E_b} \right)^{-(\alpha + \beta {\rm log}(E/E_b))} 
\end{equation}
with parameters $N_0 = 0.54 \pm 0.08$, $\alpha = 1.5 \pm 0.4$, $\beta = 0.4 \pm 0.1$, and $E_b = 2.3 \pm 0.6$.

\begin{table*}
\centering
\caption{95\% flux ($F_{95}$), energy flux($G_{95}$) , and $\nu F_{\nu}$ upper limits of HESS J0632+057 calculated over the LAT energy range.  Greyed out values indicate flux upper limits calculated over a restricted set of orbital phases that correspond to the peak emission in the X-ray \& TeV energy bands. }
\begin{tabular}{@{}c|c|c|c|c|c|c@{}}
\hline
Energy range  &  TS & \multicolumn{2}{c|}{$F_{95}$} & \multicolumn{2}{c|}{$G_{95}$} & $\nu F_{\nu}$(95\%)\\ 
GeV &   & \multicolumn{2}{c|}{$10^{-9}$ ph cm$^{-2}$ s$^{-1}$} & \multicolumn{2}{c|}{$10^{-12}$ erg cm$^{-2}$ s$^{-1}$} & $10^{-12}$ erg cm$^{-2}$ s$^{-1}$ \\
\hline
0.1 -- 0.3 &  1.7 &   $<40$   & \cellcolor{Gray}$<60$ &    $<$12 & \cellcolor{Gray}$<15$ &    $<$11\\

0.3 -- 1.0 &  4.7 &   $<9.0$ & \cellcolor{Gray}$<15$ &     $<$7.7   & \cellcolor{Gray}$<12$ &    $<$6.4\\

1.0 -- 3.0 &  4.3 &   $<1.4$  & \cellcolor{Gray}$<4.1$ &    $<$3.6   & \cellcolor{Gray}$<11$ &    $<$3.3\\

3.0 -- 10  &  0.0 &   $<0.21$  & \cellcolor{Gray}$<0.69$ &    $<$1.7   & \cellcolor{Gray}$<5.7$ &    $<$1.4\\

10 -- 30   &  0.0 &   $<0.08$ & \cellcolor{Gray}$<0.60$ &    $<$2.2   & \cellcolor{Gray}$<16$ &    $<$2.0\\

30 -- 100  &  0.0 &  $<0.06$  & \cellcolor{Gray}$<0.42$&    $<$4.8   & \cellcolor{Gray}$<35$  &    $<$4.0\\

$> 0.1$    &  6.7 &   $<30$   & \cellcolor{Gray}$<80$ &     $<$30   & \cellcolor{Gray}$<90$ &  -- \\

\hline
\end{tabular}\label{ULtab}
\end{table*}

\section{HESS J0632+057}
At no point in the analysis did the source that was added at the position of \hess\ have a significant TS value.  Using the final source model, fitting for a source at the position of \hess\ yields a TS of 4.7, far below the significance threshold for claiming a source detection

To cover the possibility that the source has a very soft spectrum in the LAT energy band, the entire analysis, including the iterative building of a source model, was repeated using all events at energies $E>100$ MeV.  The additional events incorporated into this analysis did not yield a statistically significant detection of \hess.  The region model for an $E>100$ MeV analysis required the addition of four point sources in addition to the 2FGL catalogue sources. These four sources corresponded to four of the sources that were incorporated into the $E>300$~MeV analysis and are indicated in Table~\ref{Table1}.

\subsection{Orbital variability and flaring analysis}

In the analyses described so far \hess\ was assumed to be a persistent source and we conclude that its average $\gamma$-ray emission is below the sensitivity of \fermi-LAT.  All of the other LAT-detected $\gamma$-ray binaries show periodic orbital variability including the extreme case of the periodic transient emission seen from PSR B1259$-$63  \citep{B1259,B1259Tam}, which is only detectable at HE and VHE around periastron passage.   To investigate the hypothesis that \hess\ is only detectable during a specific phase of the orbit a further analysis was performed folding the \fermi-LAT data on the orbital period of 321 days, and setting phase 0.0 to the epoch 54857 MJD \citep{Bongiorno2011}.  The orbit was subdivided into eight equally spaced phase bins, and for each of them a maximum likelihood fit was performed freezing the spectral parameters of all nearby sources in the model.  No significant emission was found in any of the phase bins. 

%Finally, the data was searched for flares or high flux states of \hess, calculating light curves with differently binned timescales, from days to months. For none of the timescales investigated was a significant signal identified. 

HESS J0632+057 has been routinely monitored by the LAT Collaboration.
A standard analysis is performed on daily, weekly and 28 day timescales to search for any flaring behaviour from the source; on these timescales, the source has never been detected by the LAT.

\subsection{Flux upper limits}
As no significant source is identified at the location of \hess, 95\% flux upper limits are derived in different energy bins using the Bayesian method developed by \citet{Helene}. With this method the 95\% flux upper limits are found by integrating the likelihood profile (function of the source flux $F$ ) starting from $F = 0$, without any assumption on its distribution. 
To calculate the values of the likelihood, the final spectral-spatial model is used with the flux normalisation factors of all the sources with TS $< 50$ set free, while freezing all the parameters of the other sources. 
We kept fixed the spectral index of HESS J0632+057 to the best fit value found in the previous analysis, $\Gamma=2.9$, while its flux normalisation factor was allowed to vary.
Table \ref{ULtab} reports the 95\% upper limits calculated in different energy bins.  Additional upper limits were calculated over a restricted range of orbital phases (0.250, 0.375) that correspond to when the X-ray and VHE fluxes are observed to peak over the orbit. In Figure \ref{upperLimitPlot} the \fermi-LAT upper limits are plotted toghether with the detection points of MAGIC \citep{magic2012}.

The uncertainties of the \fermi-LAT effective area and of the Galactic diffuse emission are the two primary sources of systematics that can affect the results derived in our analysis. 
The former systematic effect is estimated by repeating the upper limit calculations using modified instrument response functions that bracket the ``{\tt P7SOURCE\_V6}'' effective areas. 
Specifically, they are a set of IRFs in which the effective area has been modified considering its uncertainty as a function of energy in order to maximally affect a specific spectral parameter \citep{OnOrbitCal}.
The latter systematic effect is evaluated by substituting in the spectral-spatial model alternative models for the Galactic diffuse emission.  Eight different models have been generated that vary the cosmic-ray source distribution, the halo size, and the H\,\textsc{I} spin temperature.  In contrast to the standard model, the alternative models also have free independent normalisation coefficients for the inverse Compton component, and for the components associated with H\,\textsc{I} and CO, which are subdivided in four independent Galactocentric annuli \citep{SNRproc}. 
The effect of the systematic is included in the upper limits reported in Table \ref{ULtab}.

\section{Radio observations}

We tested the hypothesis that the HESS J0632+057 system is a binary in
which one of the two stars is an observable radio pulsar.  
With this aim, we searched for radio pulsations using the GBT.
%
% No periodic signals or individual dispersed bursts of a likely
% astronomical origin were detected.  As with any pulsar search, such
% observations can only constrain the brightness of a possible radio
% pulsar under the assumption that the pulsar is beamed toward us and
% not significantly scattered by the intervening interstellar medium.
%For the case of HESS J0632+057, it is also possible that the pulsar emission could be eclipsed by the wind of a massive companion, especially near periastron, in analogy with PSR B1259$-$63 \citep{Johnston1992}.
%In fact, the projected orbital separation of the putative pulsar and its companion is never larger than the projected periastron distance in the PSR B1259$-$63 system.
In order to mitigate the possible effects of scattering and eclipsing
by the wind of a massive companion,
especially near periastron, in analogy with PSR B1259$-$63 \citep{Johnston1992},
our strategy was to observe at the relatively high radio
frequencies of 2 and 9 GHz. 
Since no orbital ephemeris was available when planning these observations, 
to try to not observe the system
close to periastron, when obscuration and Doppler smearing of the
radio pulses (due to orbital motion) are at their maximum, we made a
series of 6 observations spread over the 4-month period of 24 Dec 2008
- 19 April 2009 (see Table \ref{GBTtab}).
All observations were performed with
the GBT Pulsar Spigot \citep{Kaplan2005}.  
The center frequencies were 1.95/8.90\,GHz, with 768/1024 spectral
channels spanning a bandwidth of 600/800\,MHz.  For both observing
frequencies, the sampling time was 81.92\,$\mu$s.  The maximum
dispersion measure (DM) along this line-of-sight, according to the
NE2001 Galactic free electron density model of \cite{CL2002}, is
182\,pc cm$^{-3}$.  For reference, for a moderate distance of 3\,kpc,
the NE2001 model predicts a DM of $\sim 100$\,pc cm$^{-3}$.  Given the
high spectral and time resolution of these data, the high observing
frequencies, and the relatively low expected DM along this
line-of-sight, our searches were not significantly hampered by
intra-channel dispersive smearing, even down to millisecond pulse
periods.

We used both a Fourier-based acceleration search
\citep{Ransom2002} and a search for single dispersed pulses using
the PRESTO suite of pulsar software.  We used standard search
procedures \citep[e.g.][]{Hessels2007} to clean the data
of radio frequency interference (RFI) and to dedisperse the data at a
range of DMs from $0-364$\,pc cm$^{-3}$ (up to twice the maximum
expected total Galactic column density along this line of sight).

%No periodic signals or individual dispersed bursts of a likely astronomical origin were detected.  
% As with any pulsar search, such
% observations can only constrain the brightness of a possible radio
% pulsar under the assumption that the pulsar is beamed toward us and
% not significantly scattered by the intervening interstellar medium.

Inspection of the resulting pulsar candidates revealed no signals of
likely astronomical origin.
%
% With the caveat that a putative pulsar
% may only be visible at certain orbital phases, and the fact that no
% orbital ephemeris was available when planning these observations, 
%
Under the assumption that the pulsar is beamed toward us and
it is not significantly scattered by the intervening interstellar medium,
we can place a simple flux density limit at our two observing frequencies
using the modified radiometer equation \citep{Dewey1985}.
At 2/9\,GHz we estimate a maximum flux density of
$< 90/40\,\mu$Jy (for $P_{\rm spin} >$ 1\,ms, 1-hr observation time, an
assumed pulse width of 10\%, and total system plus sky temperature of
40\,K).  This means that if the system does contain a pulsar, it must
be significantly dimmer than the bulk of the known pulsar population,
not beamed toward Earth, very distant, {\it and/or} significantly
obscured by the wind from its companion star.
%In the future, observations specifically near apastron will maximize the chances of
%detecting radio pulsations.
%, though, as mentioned above, the projected
%orbital separation is small compared with the PSR B1259$-$63 system,
%in which the radio pulsations become obscurred near periastron.

\subsection{Numerical simulations}

Within the hypothesis that the compact object is a pulsar with its radio beam pointing toward us, we performed numerical simulations of observations at both 1.95 and 8.90 GHz in order to understand at which orbital phases the pulsar could be most likely visible, and we compared results with the GBT data. In particular, we calculated the expected observed flux fraction of the pulsar radiation as a function of the orbital phase, under the hypothesis of Free-Free absorption by the strong wind spilling off the companion.

The wind of a Be star has a polar and an equatorial component.
We assumed as main features of the two components those proposed by \cite{Bogomazov} for the case of the Be star SS 2883 (companion of the binary pulsar PSR B1259-63). 
The polar wind is an isotropic component whose electron density $n_e$ follows an inverse square law $n_e(r) = n_{0e}(r_0 / r)^2$, 
where $r_0$ is the stellar radius, $r$ is the radial coordinate from the center of the star, and $n_{0e} = 10^9$ cm$^{-3}$.
The equatorial disk component is confined within an angle $\theta=\pm 7.5^{\circ}$. 
The electron density was modeled by a power law $n_e(r) = n_{0e}(r_0 / r)^{\beta}$, 
with $\beta = 2.55$, and $n_{0e} = 10^{12}$ cm$^{-3}$.
The disk was either considered as indefinitely extended or as being truncated at a radius between the periastron and the semi-latus rectum of the orbit.
Both polar and disk components are constituted by isothermal fully ionized gas with a temperature of 10000 K, i.e. the same as the surface temperature set for the star.

% We followed the best-fit model and relative parameters used in \cite{Bogomazov} for the case of SS 2883 (companion of the binary pulsar PSR B1259-63), in which the wind was modeled as a fully ionized gas constituted by two components:  an isotropic (polar wind) component supposed to extend to infinity, and an 
% equatorial (disk) component, which was either considered as indefinitely extended or as being truncated at a radius between the periastron and the semi-latus rectum of the orbit.

The results of the simulations show that in the case of a truncated disk, the pulsar would be visible, at both frequencies, over most of the orbit for almost all \footnote{Only a narrow region of the parameter space is consistent with a net decrease in the observed flux at the orbital phases corresponding to some of the non-detections. However this only occurs at the lower frequency (1.95 GHz), assuming very high inclination angles ($i \sim 90^{\circ}$) and for an increased electron density of the isotropic wind with respect to \cite{Bogomazov} best-fit model.} the explored geometrical configurations of the system; in particular, the GBT non-detections would always fall at orbital phases at which we would expect to observe 100\% of the total flux or so, thus indicating that the hypothesis of a truncated disk for this system is not compatible with the present data.
In the case of an extended disk, none of the pulsar orbital phases can be selected as being safe for a possible pulsar detection.
Largely because of our lack of information on  the disk orientation and physical characteristics, we were always able to find a proper combination of geometrical parameters to put the eclipse interval at any possible orbital phase. If the pulsar signal were finally detected we would then be able to put significant constraints on the system geometry.

\begin{table}
\centering
\caption{Summary of GBT Observations. Orbital phases are calculated with respect to the phase zero as defined 
by Casares et al.(2012) (at MJD = 54857.0) and using an orbital period
of 321 days.  For reference, periastron and apastron occur at phases 
0.967 and 0.467, respectively}
\begin{tabular}{@{}ccccc@{}}
\hline

Date        &    MJD     &   Obs. Time  &  Obs. Freq.  & Orb. Phase  \\
            &    (Start) &      (s)     &  (GHz)       &    \\
\hline
24-12-2008   &   54824.140 &  3000      &  2    &   0.90  \\
23-03-2009   &   54913.853 &  5000      &  9    &   0.18\\
26-03-2009   &   54916.863 &  5000      &  2    &   0.19\\
27-03-2009   &   54917.947 &  3900      &  2    &   0.19\\
30-03-2009   &   54920.814 &  5000      &  9    &   0.20\\
19-04-2009   &   54940.832 &  8500      &  2    &   0.26\\
\hline
\end{tabular}\label{GBTtab}
\end{table}

\begin{figure}
\center
\includegraphics[width=0.45\textwidth]{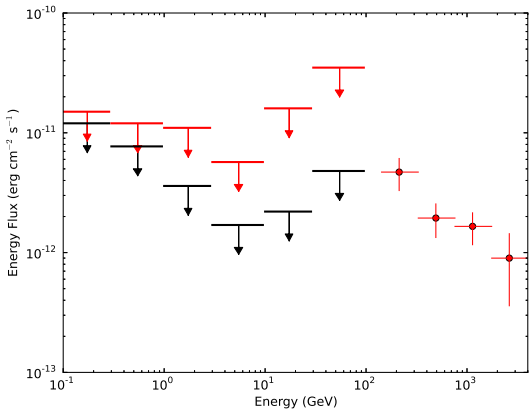}
\caption{The high energy SED of HESS J0632+057 showing the 95\% energy flux upper limits calculated from the LAT data using all phases (\textit{in black}). Are also shown the 95\% energy flux upper limits calculated using orbital phases which correspond to the X-ray peak (\textit{in red}), together with the MAGIC measurements (\textit{red data points}) corresponding to the source detection in the same orbital phases \citep{magic2012}.}
\label{upperLimitPlot}
\end{figure}

\section{Discussion}

A detailed analysis of $\sim$3.5 years of \fermi\ LAT data does not yield a significant detection of the $\gamma$-ray binary \hess. Similarly, searches for persistent, orbitally variable or flaring emission did not yield any detection. This makes this object unique among known $\gamma$-ray binaries as the only member of the class
eluding detection in the 0.1--100 GeV energy band, with an energy flux upper limit of $<3 \times 10^{-11} {\rm erg} ~{\rm cm}^{-2}~ {\rm s}^{-1}$.  

Radio observations were also conducted with GBT. 
No periodic signals or individual dispersed bursts of a likely astronomical origin were detected.
We estimated the flux density limit of $< 90\,\mu$Jy and $< 40\,\mu$Jy at frequencies of 2\,GHz and 9\,GHz, respectively, assuming the pulsar is beamed toward Earth.

Some of the properties of the confirmed $\gamma$-ray binaries are summarised in Table~\ref{binary_pop}.  Upon first glance the multi-wavelength properties of \hess\  exhibit similarities to PSR B1259$-$63 and LS~I~+61$^\circ$303:  
\begin{itemize}
\item[-] All three systems have Be stars as the companion star.
\item[-] The \hess\ orbital period lies in the range between LS~I~+61$^\circ$303 and PSR B1259$-$63.
\item[-] The VHE and X-ray emission are in phase in both \hess\ and LS~I~+61$^\circ$303.
\end{itemize}

\noindent LS~I~+61$^\circ$303 is one of the brightest persistent GeV sources detected by the LAT and is detectable throughout its orbit while PSR B1259$-$63 
is characterised by transient $\gamma$-ray emission, 
as evidenced by the LAT detecting the source only around periastron passage.
Of all the known $\gamma$-ray binaries 
\hess\ has the closest estimated distance to Earth and consequently it was anticipated as a strong candidate for detection at GeV energies by \fermi.

%NEW PARAGRAPH
The non-detection is all the more intriguing in light of the recent report by \citet{magic2012} who detected emission from \hess\ with the MAGIC telescope from 136 GeV -- 4 TeV in 2011 February.
They reported extending the spectral detection of the source into the 136--400 GeV band, and their spectral fit to the data indicated a pure power law shape, consistent with the spectrum reported by the H.E.S.S. collaboration \citep{hess_discovery}, with no sign of a spectral turnover. The LAT upper limits calculated in the same phase range as the MAGIC observations 
are plotted in Figure~\ref{upperLimitPlot} in red, with the MAGIC measurements in blue. 
The LAT 95\% flux upper limits are sufficiently constraining to indicate that the VHE spectrum observed by MAGIC and HESS cannot be extrapolated back by a simple power law to the LAT energy band and, consequently, there must be a turnover in the spectrum between 10--136 GeV in order for the source to not be detectable in the available LAT dataset. 
The LAT upper limits calculated over all orbital phases are plotted in black in Figure~\ref{upperLimitPlot}.
With the current results we can not exclude that at HE the orbital flux modulation of \hess\ is a negligible fraction of its overall flux. Under this hypothesis the comparison of the LAT upper limits calculated over all the orbit with the MAGIC measurements give a more stringent constraint of the spectral turnover, which should happen close to 100 GeV.

In the context of the well-studied $\gamma$-ray binaries, LS~I~+61$^\circ$303 and LS~5039, a disconnection between the HE and VHE emission components is not entirely unexpected.  The LAT spectra of LS~I~+61$^\circ$303 and LS~5039 are best described by a power law with an exponential cutoff that does not extrapolate up to the reported VHE spectra \citep{2012ApJ...749...54H}.  Both sources have also shown indications of a high-energy tail that may connect to the VHE emission. However it is clear that the GeV emission is dominated by the cutoff power law spectral component which has led to the suggestion that two different emission processes may be in action to produce the HE and VHE emission detected in these binary systems.

\subsection{Striped wind emission from a pulsar?}
A potential explanation for the ``missing'' HE emission from \hess\ has been suggested by \citet{2011MNRAS.417..532P}.  They propose that the HE emission detected from PSR B1259$-$63
%, the only $\gamma$-ray binary in which the nature of the compact object is known, 
can be explained by inverse Compton up scattering of stellar photons by particles within a ``striped pulsar wind''.  In this model the HE emission is produced in the vicinity of the pulsar, explaining the ``pulsar-like'' exponentially cutoff power law spectrum observed by the LAT while the radio/X-ray/VHE emission is produced beyond the termination shock of the pulsar wind.  Such a model explains the disconnect between the HE and VHE emission in the $\gamma$-ray binaries.  For the HE emission to be detectable the Earth line-of-sight must pass through the ``striped wind''. Assuming the pulsar obliquities are isotropically distributed then this would only occur in 50\% of $\gamma$-ray binaries.  

\begin{table*}\caption{A comparison of the general high-energy properties of the confirmed $\gamma$-ray HMXBs.  Persistent HE sources are quoted with mean fluxes in the LAT energy band while transient sources are specified by a flux range.}
\begin{threeparttable}[b]
\begin{center}
\begin{tabular}{lcccccccc}
\hline
Source & Companion & Distance & Orbital period & Persistent/ & F$_{0.1-100\,{\rm GeV}}$  \\
 & sp. type & kpc & days & Transient & $10^{-10}\,{\rm erg}\,{\rm cm}^{-2}\, {\rm s}^{-1}$ \\
\hline
HESS J0632+057\tnote{a} & B0Vpe & 1.4 & 321 & -- & $<0.3$\\
PSR B1259$-$63\tnote{b} & O9.5Ve & 2.3 & 1236.79 & Transient & (0.9--4.4)$^{\ddag}$$^{g}$\\
LS~I~+61$^\circ$303\tnote{c} & B0Ve & 2 & 26.496 & Persistent & 5.0$^h$   \\
LS~5039\tnote{d} & O6.5V((f)) & 2.5 & 3.90603 & Persistent & 2.9$^h$\\
1FGL J1018.6$-$5856\tnote{e,f} & O6V((f)) & 5.4 & 16.58 & Persistent & $2.8$$^e$\\
%Cyg X-3\tnote{f} & WR & 7 & 0.2 & Transient &  (7.6--16.8)$^{\ddag}$$^j$ \\

\hline
\end{tabular}
\begin{tablenotes}
\item $^{\ddag}$ Range given due to the transient/flaring nature of the source.\\
$^a$\,\cite{casares2012}; 
$^b$\,\cite{2011ApJ...732L..11N}; 
$^c$\,\cite{1998MNRAS.297L...5S};  
$^d$\,\cite{2005MNRAS.364..899C};    
$^e$\,\cite{1FGLbinary_science}; $^f$\,\cite{2011PASP..123.1262N}; 
$^g$\,\cite{B1259}; $^h$\,\cite{2012ApJ...749...54H};
%$^f$\,\cite{2009Sci...326.1512F};  $^j$\,\cite{2012MNRAS.421.2947C}
\end{tablenotes}
\end{center}
\label{binary_pop}
\end{threeparttable}
\end{table*}%

\section{Conclusions}
A rigorous search for emission in the GeV energy band from the $\gamma$-ray binary \hess\ using $\sim$3.5 years of \fermi-LAT data has not yielded a significant detection.  There is no significant persistent, flaring or orbital phase-dependent emission.  This makes \hess\ unique among the handful of known $\gamma$-ray binaries as the only source eluding detection by the LAT.
%The prospect of future detection with more data is not encouraging unless the source exhibits significant variability.

The 95\% LAT flux upper limit for the 0.1--100 GeV energy band of $<3 \times 10^{-8} {\rm cm}^{-2}~ {\rm s}^{-1}$ is consistent with the flux predicted by the one-zone SSC model shown in Figure 4 of \citet{hinton2009} and \citet{skilton2009},
which assume a magnetic field of B = 70~mG and a radiation density of U$_{rad}$= 1~erg~cm$^{-3}$.
%The 95\% LAT flux upper limit for the 0.1--100 GeV energy band of $<3 \times 10^{-8} {\rm cm}^{-2}~ {\rm s}^{-1}$ is a strong constraint for any models invoked to explain the HE and VHE emission of this object.  
However, it is clear from the LAT data that there must be a turnover in the VHE spectrum of \hess\ below $\sim$136 GeV in order for the \fermi-LAT data to be compatible 
with the data obtained at higher energies by MAGIC and H.E.S.S.
The ``striped pulsar wind'' model invoked by \citet{2011MNRAS.417..532P} to explain the emission from PSR B1259$-$63 could explain what is observed in \hess, however a pulsar nature for the compact object in this system has not been confirmed and hence other models may offer feasible explanations.

Numerical simulations were performed to calculate 
the radio flux fraction expected to be observed along the orbit by a pulsar with its radio beam pointing toward us, while absorbed by Free-Free interaction with the strong wind (polar and disk) from the companion.
The results of the simulations lead us
to exclude the truncation of the disk in favor of
an indefinitely extended disk. Alternatively we could
exclude the hypothesis that the radio beam of the
plausible pulsar is pointing toward the Earth.

\section*{Acknowledgments}

The \textit{Fermi} LAT Collaboration acknowledges generous ongoing support
from a number of agencies and institutes that have supported both the
development and the operation of the LAT as well as scientific data analysis.
These include the National Aeronautics and Space Administration and the
Department of Energy in the United States, the Commissariat \`a l'Energie Atomique
and the Centre National de la Recherche Scientifique / Institut National de Physique
Nucl\'eaire et de Physique des Particules in France, the Agenzia Spaziale Italiana
and the Istituto Nazionale di Fisica Nucleare in Italy, the Ministry of Education,
Culture, Sports, Science and Technology (MEXT), High Energy Accelerator Research
Organization (KEK) and Japan Aerospace Exploration Agency (JAXA) in Japan, and
the K.~A.~Wallenberg Foundation, the Swedish Research Council and the
Swedish National Space Board in Sweden.
Additional support for science analysis during the operations phase is gratefully
acknowledged from the Istituto Nazionale di Astrofisica in Italy and the Centre National d'\'Etudes Spatiales in France.

Work done in the framework of the grants AYA2009-07391, AYA2012-39303,
as well as SGR2009- 811, TW2010005 and iLINK2011-0303.

A.~B. Hill is supported by a Marie Curie International Outgoing Fellowship within the 7th European Community Framework Programme (FP7/2007--2013) under grant agreement no. 275861. 

G. Dubus is supported by the 7th European Community Framework Programme (FP7/2007--2013) under grant agreement no. ERC-StG-200911.

P.~H.~T. Tam is supported by the National Science Council of the Republic of China (Taiwan) through grant NSC101-2112-M-007-022-MY3.

\label{lastpage}

\end{document}